# Public Computing Intellectuals in the Age of AI Crisis

A Preprint

**Randy Connolly**

Department of Mathematics & Computing

Mount Royal University

Calgary, Canada

rconnolly@mtroyal.ca

## Abstract

The belief that AI technology is on the cusp of causing a generalized social crisis became a popular one in 2023. While there was no doubt an element of hype and exaggeration to some of these accounts, they do reflect the fact that there are troubling ramifications to this technology stack. This conjunction of shared concerns about social, political, and personal futures presaged by current developments in artificial intelligence presents the academic discipline of computing with a renewed opportunity for self-examination and reconfiguration. This position paper endeavors to do so in four sections. The first explores what is at stake for computing in the narrative of an AI crisis. The second articulates possible educational responses to this crisis and advocates for a broader analytic focus on power relations. The third section presents a novel characterization of academic computing's field of practice, one which includes not only the discipline's usual instrumental forms of practice but reflexive practice as well. This reflexive dimension integrates both the critical and public functions of the discipline as equal intellectual partners and a necessary component of any contemporary academic field. The final section will advocate for a conceptual archetype— the Public Computer Intellectual and its less conspicuous but still essential cousin, the (Almost) Public Computer Intellectual—as a way of practically imagining the expanded possibilities of academic practice in our discipline, one that provides both self-critique and an outward-facing orientation towards the public good. It will argue that the computer education research community can play a vital role in this regard. Recommendations for pedagogical change within computing to develop more reflexive capabilities are also provided.

## 1 Introduction

> "The crisis consists precisely in the fact that the old is dying and the new cannot be born; in this interregnum a great variety of morbid symptoms appear."
>
> —Antonio Gramsci, *Prison Notebooks* (1930) [155]

The perception that we are currently living in "an era of escalating, overlapping crisis" [69] is maybe one of the few statements that can generate wide-scale agreement even in our opinion-polarized times [126]. Despite several decades of immanence, the ecological crisis reminds us anew of its reality almost every summer. Within the United States, the George Floyd murder brought increased attention to the persistent crisis of continuing state violence against racial minorities. Financial crisis may have the oldest pedigree, rearing itself every few decades over the past 150 years; but the economic slowdown of the 2000s along with growing inequalities in wealth have resulted in an almost perpetual state of economic crisis [86]. Since 2016, a conviction has grown that many states in the developed world are experiencing a legitimation crisis in which citizens are losing their trust in the key institutions of their society [92]. The covid pandemic of 2020-2023 was arguably the most acute and universal of crises, both in its global scope and in terms of its impact on individual and collective life. And now with Large-Language Models (LLM) and generative AI[1] algorithms, we appear to have a new and unwelcome guest in the lineup of undesirables that constitute our contemporary procession of crises [119, 72].

Technological change inevitably brings anxiety with it. Concern about the implications of computerization has been a persistent shadow that has accompanied the computer right from the very first public reveals of the ENIAC in 1946 [118]. Sometimes the shadow has been barely visible, for instance, during periods of enthusiasm that accompanied new technological breakthroughs, such as time-sharing (early 1960s), the PC (early 1980s), the web (mid 1990s), and social networking (late 2000s) [33]. But at other times, this shadow of concern has almost eclipsed the technology itself. Much of this computer imaginary—that is, the discourse of fear and hope around the computer—came from outside computing [11]; nonetheless, there has always been a shadow discourse within computing that has also been wary about computing advances. Norbert Weiner, whose early pioneering work on cybernetics and automation at MIT played an important role in the development of computer science as a research area, stressed in his best-selling 1950 *The Human Use of Human Beings*, that human control is vital for a technology whose inner logic is specifically oriented towards removing human control [82]. Joseph Weizenbaum, a computing colleague of Weiner at MIT, also gained public prominence with his 1976 *Computer Power and Human Reason*, which articulated a set of critiques about the newly-emerging field of artificial intelligence [46]. In the 1980s, these concerns were institutionalized within computing with the emergence of computer ethics and society studies as a small but vibrant part of the computing discipline itself. Concerns articulated in early textbooks such as Johnson's *Computer Ethics* [96] and the Dunlop and Kling's *Computerization and Controversy* [56] were eventually echoed by the ACM and IEEE in codes of conduct [4] and curricular recommendations.

While social, political, and psychological concerns about computing advances are by no means new, recent advances in AI have foregrounded social concerns in a very public fashion that do seem quantitatively different than previous moral panics about technological change. Why has this been the case? Certainly, we are often told that AI innovation is uniquely different from all previous automation technologies [e.g., 24, 91, 164]. That is, one of the key aspects of AI lies in the capacity of these systems to operate somewhat autonomously through the recognition of unseen patterns; as such, it seems to threaten a wider range of cognitive work in comparison to previous forms of computerization. These concerns have no doubt been heightened by the fact that some of the most prominent voices of worry are coming from the tech sector itself. "Top AI researchers and CEOs warn against 'risk of extinction'" [176] and "Elon Musk among experts urging a halt to AI training" [173] were just two of the eye-catching headlines in the popular media about AI in 2023. Several high-profile executives of AI enterprises, such as Sam Altman, CEO of OpenAI and Igor Babuschkin of Elon Musk's xAI, unexpectedly joined the chorus of anxiety [31]. It is worth noting that these high-profile statements of worry about our future artificial general intelligence (AGI) overlords are focused not on the present-day capabilities of this technology, but on hypothetical far-future ones. One reason for this might be the burgeoning interest in so-called longtermism [168] and Effective Altruism [74] amongst the executive class of Silicon Valley. Or it is possible that these statements are simply a kind of AI crisis theatre, where executives make grand pronouncements about the catastrophic risk potential of all-powerful AI, all the while continuing to use/sell/market AI technology. If so, such pronouncements are yet another installment of high-tech hype, overselling capabilities in public in order to better sell it in private [11]. But as noted by a variety of commentators [100, 45, 38, 122], these declarations of future risk are also prevarications: they deflect attention away from the already-existing version of AI that is already causing harm. Haunting futures of the soon-to-be (or maybe-already-here)—which include mass unemployment [70, 166], deskilled labor pools [3, 50], feudalized political economies [131,75,174], and de-autonomized subjects dominated by algorithmic governance [187, 101]—have sparked another resurgence of interest in the consequences of computing technology not only in the world at large but also, as this essay will later illustrate, within the more pragmatic hallways of academic computing departments.

---

[1] As noted by Crawford [45], AI is neither artificial, nor intelligent. It is, in the words of Lindgren [115], an empty signifier. This means that "AI can mean many different things, not only in technological terms but, more importantly, as regards what political meanings are symbolically connected to it, and what consequences this gets in terms of both ideology and social practice" [115]. Nonetheless, the term "artificial intelligence" has become widely used to refer to a wide-range of broadly-related algorithmic approaches. Computer scientists often instead use more specific terms such as machine learning (ML), neural networks, categorisation systems, natural-language processing (NLP), or the specific algorithm being used and evaluated.

In the Gramsci quote that begins this essay, the morbid symptoms were the fascist regimes that sprouted across central Europe in response to the economic and legitimation crises of the 1920s and 1930s. In my essay, the morbid symptoms refer to something much less dramatic, namely a general tendency within academic computing to downplay the social and political ramifications of our work. One possible reaction to crisis is to insist there is no crisis, an attitude popularly captured in the "This is Fine" meme, in which a smiling hat-wearing dog sits with his coffee in a kitchen engulfed in flames, while complacently uttering "This is Fine". Yes, as noted above, the shadow discourse of concern within computing have been an important part of computing from its earliest days. Yet, if one looks at curricular recommendations[2], or at the entirety of the IEEE and ACM libraries, or at the oft-expressed frustrations amongst those teaching computer ethics [66], expressing or evaluating social and ethical concerns have been a comparatively small part of "normal" practice in academic computing. There are other proxies that illustrate this fact. For instance, as noted by Birhane *et al* [17], there heretofore has been a noticeable lack of interest in the wider ramifications of AI technology by practicing ML researchers: only 1% of the most highly-cited ML papers even mentioned possible social and ethical concerns much less analyzed them. Similarly, Santy *et al* [151] found an almost complete absence of formal Ethics Review Board submissions by NLP researchers across the 2010s and early 2020s. Even those ML papers that explicitly focus on problems of bias often do so without any form of normative reasoning or social theory, but instead treat it strictly as a technical problem that has a technical solution [19]. Of course, within the discipline of computing, we may feel considerably less worry about these technologies, partly because we are more likely to be aware that they are elegant predictive applications of statistics and linear algebra and nothing remotely close to AGI, partly because we may believe that there are possible socially-beneficial outcomes with this technology [47, 24], and partly because we as a discipline stand to gain the intellectual capital of prestige or the financial capital of ever-increasing grants as this technology progresses. But we inhabit the same world as everyone else: our disciplinary expertise won't shield us from (possible) wide-scale social upheaval.

If ignoring a crisis is unwise, we also need to recognize that the rhetoric of crisis is simultaneously both a moment of danger and an opportunity for transformation [169]. Whether the AI crisis, like any of the other crises mention in the opening sentences of this paper, is real or imagined or somewhere in between doesn't really matter for the purposes of this paper. My focus in this paper is on crisis as a potential opportunity, a critical moment that allows the academic discipline of computing to question the trajectory of its current path. For this reason, crisis and critique are intrinsically linked: crisis prompts critique of the social conditions that induced it, and critique apprehends the reality of the crisis for all to see [43]. It is this essay's contention that this relatively rare conjunction of shared public concern about social, political, and personal futures presaged by current developments in artificial intelligence not only among the wider community but also within computing itself, presents the academic discipline of computing with a rare opportunity "to interrogate the normativity currently in place and take actions for change" [190]. As a position paper, this essay will use the AI crisis (real or imagined) to make an argument for an expanded conception of academic computing and the role of both individual computer scientists and computer education research in that transformation.

---

[2] It should be noted that in the most recent ACM 2023 CS Curricular Recommendation, there is much more emphasis on "Society, Ethics, and the Profession" topics, with almost twice as many hours devoted to these topics in comparison to the previous 2013 CS Curricular Recommendation. But even so, even with this doubling in the latest recommendation, such topics are only 6.7% of the total recommended CS Core hours.

This essay's first section expands on the nature of the AI crisis for computing. It will argue that the problems with this technology lie not in the realm of philosophy, addressable straightforwardly through ethics or codes of conduct, but in the social realm, and thus are best interrogated with the lenses of economics, political science, and sociology. The second section will look at educational responses to the AI crisis; in particular, it will focus on approaches that advocate addressing the relations of power that wrap any technological artefact. The third section presents a novel categorization of academic computing's field of practice, one which includes not only the discipline's usual instrumental forms of practice but reflexive practice as well. This reflexive dimension integrates both the critical and public functions of the discipline as equal intellectual partners and a necessary component of any contemporary academic field. The final section will advocate for a conceptual archetype—the Public Computer Intellectual—as a way of practically imagining the expanded possibilities of academic practice in our discipline, one that provides both internal critique and an outward-facing orientation towards the public good. It will argue that the computer education research community can play a role in this regard.

## 2   AI: An ethical or political crisis?

> "In 2023 everything speaks AI. The AI gaze is present in every public discourse and the AI question is present in every interview."
> —Londi Ziko [191]

During the first few months of 2023, it was difficult not to read, watch, or hear both acclaim and apprehensions about the latest manifestations of artificial intelligence. Within the university context, almost everyone, whether students, faculty, or administrators, had some type of conversation about Large-Language Models (LLM) / generative AI and their potential impact on the way we teach and students learn. ChatGPT was the most visible of these products, allowing students to effortlessly produce standard university assessments such as the five-paragraph summary paper about a well-known text or a recursive bubble sort in C++. The GitHub Copilot extension for the Visual Code text editor is an especially powerful use of specialized LLMs that seemingly eliminates a lot of the cognitive friction involved in solving common problems that are part of learning how to program. Popular text-to-image generators using diffusion models (such as Midjourney and DAL-E) have given rise to existential dread (and justified outrage over the theft of their intellectual property) for anyone who relies (or hopes to rely) on their ability to create or manipulate visual assets for a living. Similar products for generating music [89] and video [88] appear to be imminent. The pace of change in this area has accelerated to such an extent that conference proceedings are no longer agile enough. Instead, the non-peer reviewed online pre-print service `arxiv.org` has become the dissemination vehicle of choice within the fervid AI research community.

Over the past decade, the general topic of AI has been of growing interest to philosophers, economists, sociologists, and political scientists. Concerns over bias and discrimination in data science in general, and in automated decision-making systems in particular, have been widely acknowledged [137, 136, 62, 14, 107, 128, 25, 19]. At the same time, within academic computing there emerged a consensus that ethics instruction should be the main mechanism for instilling social accountability into the field [4, 66, 36, 162].[3] This embrace of ethics has been especially ubiquitous within AI research communities [36, 150]. Back in 2019, one study found 84 different AI ethics initiatives that had been promulgated by academic, government, and industry bodies [95]. Some have dismissed this effort as a form of 'ethics theatre' or 'ethics washing' in which the professed public statements of ethical self-commitments are mainly a show for outside observers that is ultimately motivated by the industry's desire to prevent or delay legal regulations [15, 67, 134,189]. Nonetheless, governments have belatedly become more interested in considering the ethics of AI—for instance, the EU's High-Level Expert Group on Artificial Intelligence (HLEG-AI) published the high-profile "Ethics Guidelines for Trustworthy AI" in 2019 [160]. Yet the vast majority of the members of the HLEG-AI were from industry with only a small number of ethics experts; as a result, the HLEG-AI is arguably principally interested in advancing the global standing and business opportunities of European corporations working in AI [85]. The AI crisis of 2023 (again, real or imagined) does, however, seem to have made some governments more willing to countenance regulatory approaches to AI. In March 2024, the European Parliament passed the Artificial Intelligence Act, which includes bans on a variety of intrusive and discriminatory uses of AI, regulatory obligations for providers of high-risk systems, and mandatory regulatory assessments of all AI systems [63]. A similar act passed its second reading in the Canadian Parliament in late 2023 [30]. The US has been slower to act in this area. But in July 2023, the Biden-Harris administration announced that they had secured "voluntary commitments" from a subset of tech companies to practice "safety, security, and trust" as a purported first step in developing future legislation around AI technology [181].

This move towards regulation is perhaps tacit recognition that articulating general ethical guidelines, while admirable in principle, often lack actionability in that these codes "do not offer specific practices to apply ethics at each stage of the AI/ML pipeline and often fail to be actioned in governmental policy" [36]. Furthermore, recent studies have found that knowledge of ethical guidelines and/or codes of ethical conduct do not seem to affect the decision making of software developers [121,172,183], a finding which accords with similar studies evaluating the effectiveness of business ethics codes [105]. For this reason, some commentators have asserted that "AI ethical principles are useless, failing to mitigate the racial, social, and environmental damages of AI technologies in any meaningful sense" [132; see also, 134, 188, 129]. One of the leading voices in computer ethics has similarly concluded that "we should not yet celebrate consensus around high-level [ethical] principles that hide deep political and normative disagreement" [127]. Instead, practicing computer scientists and computer educators have been urged to embrace the "difficult work which requires engaging with social and political questions" [132].

This mention of political questions is important, as it reflects a burgeoning interest not just in the normative[4] aspects of computer systems but their political aspects as well. The "fundamental truth buried in the language of statistics and computer scientists" claims Simons [159], research scientist on Facebook's Responsible AI team, is that "machine learning is political" since such systems inevitably prioritize the material interests of some social groups over others. This inequality of distribution when it comes to the benefits and burdens of computing innovation is eliciting more attention by computing academics and professionals [85, 10, 130, 79, 124, 182]. The classic text advocating this line of thinking is Winner's 1986 "Do Artifacts Have Politics?" [184]. Winner argued that technological systems can become political when they prevent certain social outcomes within a community (such as Robert Moses designing the height of New York freeway underpasses in the 1950s to prevent buses carrying lower income racial minorities from accessing public beaches) or when

---

they privilege specific social outcomes (Winner's example here was how the mechanical tomato harvester encouraged the transformation of the tomato business away from small family farms to one dominated by large agribusinesses).

The example of Robert Moses's discriminatory bridges is especially relevant here. As noted by Paltieli [138], the implementation of this bridge technology "was unquestionably political, but it also eliminated politics." That is, to have achieved the same outcome as the bridge technology (i.e., preventing lower-income racial minorities from reaching a public beach) using non-technical means would have necessitated political buy-in from the civic government, the crafting of bylaws, the involvement of police, and likely challenges by the media and the courts. The key point here is that something political happened, but thanks to technology, it happened in a way that was hidden from the public because it eliminated discussion, explanation, and debate, which are the bedrock of democratic politics [38].

Thus, rather than examining a computing practice with abstract ethical principles, we would do better, in the words of science fiction writer and technology critic Cory Doctorow, to "stop thinking about what technology does and start thinking about who technology does it to and who it does it for" [54]. While Winner's paper is the classic expression of this type of thinking, there are more and more examples of scholars doing the same work today within computing. Benjamin's *Race After Technology* [14] and Buolamwini's *Unmasking AI* [26] reveal the many ways that software developers encode biased judgments into their technological systems (a phenomenon Buolamwini evocatively called the 'coded gaze') that are just as discriminating as Moses's bridges (see also 113, 102, 182, 99). Similarly, Winner's concerns about how tech systems can privilege some social actors and burden others has also been taken up, albeit more slowly, within computing in general, and in the HCI research community in particular [53, 186, 104]. The use of AI in social contexts has made this privilege-a-few/burden-many-others dynamic an emerging area of research within computing. As noted by McQuillan, "socially applied AI has a tendency to punch down: that is, the collateral damage that comes from its statistical fragility ends up hurting the less privileged" [122].

This approach has led some critics to adopt a political economy approach to thinking about digital technology innovation. That is, critics are incorporating the point made by Zuboff in her 2019 *Surveillance Capitalism*, that "in a modern capitalist society, technology was, is, and always will be, an expression of the economic objectives that direct it into action" [192]. Perhaps the most salient political economy concern revolves around employment and the nature of work. Concerns about computerization on the quantity and quality of work have been a long-standing shadow; older textbooks such as *Computerization and Controversy* [56] and *Computers, Ethics and Society* [60] feature substantial sections on this topic. It should be mentioned that in traditional economic theory, technological change is typically modelled as productive overall for employment [7]. How so? First, there is a displacement effect ($d$) as demand for human labor declines in tasks that can be automated or performed by the new technology; second, there is a productivity effect ($p$) as labor demand increases for nonautomated tasks; and third, there is a reinstatement effect ($r$) as new categories of tasks are created due to the technology. For much of the twentieth century the evidence indicated that given time (typically around a decade), $p+r>d$ [6]. But over the past 30 years, it appears that computer automation's displacement effect has grown, while the productivity and reinstatement effects "have been slower to materialize and smaller than expected" [170]. Thus, the consequence of IT innovation over the past three decades has arguably been stagnating labor demand, lower productivity growth, and rising inequality [3, 8, 7].

The worry is that AI will accelerate these trends by being adopted commercially principally as a way of reducing labor costs rather than as a way of improving labor's productivity [2]. At present, early commercial applications of this technology—such as commercial chatbots, resume filtering, and loan approvals—have very much been focused on displacement with little to no reinstatement [68, 51]. As a result, there is worry that this type of AI will "generate benefits for a narrow part of society that is already rich and politically powerful" [2], thereby exacerbating growing inequality in advanced economies [55]. Ernst [61] calls this the "AI trilemma": that due to the reliance of machine learning innovation on gargantuan data sets processed on immense energy-consuming GPU farms affordable only by the largest of technological firms, it isn't possible to have AI productivity, lower economic inequality, and ecological sustainability. Ernst argues that two of the three could be achievable. But with the current technological paradigm in AI and with the current organization of the digital economy, only the AI productivity goal is likely to be achieved in the future at the cost of ever rising inequality and environmental destruction. Adjudicating between these three conflicting goals, it should be stressed, is not a matter of ethics or economics, but of politics.

While there is a tendency to situate the potential problems of AI in the future, we can perhaps more clearly ascertain the political nature of AI if we look at already-existing AI. For instance, the investigations of Christl [35], Crawford [45], Dzieza [57], Hao, Freischlad, and Hernández [83, 84], and Perrigo [140] into worker experience in call-centers, fulfillment warehouses, AI labelling centers, and in on-demand gig platform work didn't reveal job losses, but revealed an AI-enabled intensification of work, a work that is more accelerated, more surveilled, more demanding, and more physically grueling. As noted by McQuillan, thanks to its affordances, this already-existing AI is focused completely on maximizing the extraction of value from each worker and uses that same activity to threaten their replacement [122]. That is, many already-existing AI systems in actual use mainly serve to optimize already-existing socio-economic processes, thereby perpetuating and even amplifying already-existing inequalities. It is for these reasons, then, that more researchers in the general AI ethics space are now arguing that we need to consider political issues much more frequently.

We began this section with the observation that recent advances in AI are engendering worries that we may be on the cusp of some type of societal crisis brought on by AI. Certainly, the generalized insight that this technology stack is qualitatively different in its social risk potential does appear to be well placed. Again, the risk is not of a far-future AGI, but of an already existing technology that is damaging the natural environment, poisoning our information ecosystem, and reproducing systems of oppression by further increasing the concentration of economic power [22]. Giddens [76] and Beck [13] recognized in the last century that the emergence of human-created risk is one of the key characteristics of late modernity, and that we all have to accommodate ourselves to risk and uncertainty. But the nature of a digital risk is difficult to identify, as the objects at risk (such as autonomy, meaningful work, cognitive enfeeblement) are abstractions. As a consequence, it is difficult to engage in public debate on digital risks in order to address them through governance and regulation (i.e., engage in politics) [65, 111]. In the next section, this paper will argue that computing educators—with our interest in non-technological metrics such as universalism, fairness, diversity, collaboration, justice, and critique—can play an important role in this task. By both inclination and knowledge, computing educators have a special calling, a unique responsibility to communicate and engage in an on-going debate about the social meaning of digital technology both within the discipline and with the public at large.

## 3    Educational responses to the AI crisis

> "To change what it seems impossible today to change, we have to start changing what is possible today to change."
> —Paulo Freire (quoted in [123])

The focus within academic computing has naturally been on the topics close to us, such as algorithms, computational efficiency, programming techniques, and so on. But the depth of field of computation extends far beyond these topics. As the previous section argued, one of the values of the AI crisis narrative frame is that it can help us broaden our focus and see what's at stake for those outside academic computing. As noted in the last section, for much of the previous twenty years, academic computing has hoped that ethics instruction for its students was a sufficient means for addressing the profoundly impactful nature of ongoing digital transformation. This "normative turn" identified by Abebe *et al* [1] in computer science in general and machine learning in particular has played a key role in inculcating a wider willingness to include non-technical metrics such as fairness, bias, accountability, and transparency when designing and evaluating software systems. But as noted by Selbst *et al* [156], fairness and bias are properties of social and legal systems. If we wish to prevent, for instance, discriminatory outcomes in our computing systems, as argued in the previous section, computing researchers must first engage with the social and political environments in which inequality and injustice arise [158, 73]. Despite many educators' desire for a neutral stance in the classroom divorced from the messiness of real-world politics, computing educators need to accept that this is unrealistic, and, increasingly in the future, they will need to "adopt normative positions on issues they probably prefer to avoid" [110]. Indeed, it has long been recognized within political science that arguing that one's work is unpolitical is, in fact, a very political statement, one that advocates for the conservation of the status quo [146].

A growing body of scholarship within computing education has already taken up the challenge of addressing the political aspects of our field. Vakil [171] was an early advocate for a justice-centered approach to teaching computing. Instead of focusing on ethics, such an approach, Vakil argued, requires "considering the sociocultural and sociopolitical contexts in which technologies have been developed and applied." In such a course, students would engage in the critique of abuses of technological power and learn how to design technological systems that reflect students' social and civic identities. Recent special issues on justice-centered computing education in *ACM Transactions on Computing Education* similarly called for a reorientation "of computing and computing education towards justice goals" [109].

How is this to be achieved? One key step is to broaden the academic perspective of computing to include those of the social sciences [40]. Similarly, Dignum [52] declared that AI is (or should be) by nature a multidisciplinary endeavor and that "it is essential to integrate humanities and social sciences into the conversation about AI". Indeed, focusing too closely on the near-to-hand technological aspects of AI does not allow us to see what is morally at stake with this particular technology. AI consists not only of the *technical approaches* (the algorithms, the data models), but also *industrial infrastructures* (almost limitless reservoirs of data, vast quantities of processors for running the models, subsidized energy grids) and *social practices* (protective intellectual property regimes, ready access to AI expertise, reliance on inexpensive data labeling outsourced to the global south, a lack of external regulatory policies, rent-seeking platform monopolies) [45, 175]. It is when we focus our analysis on these background social practices and industrial infrastructures that we can see the truth about AI: that like the computer industry all along—whether it be IBM in the 1950s and 1960s, Microsoft in the 1980s and 1990s, or Apple in the 2000s all of whom were grown and protected via government investment in innovation [133]—AI is a technology that is profoundly intertwined in relationships of power [38].

Over the past half century, power has been a key vector of analysis within the social sciences. Recognizing the different aspects of power can also be helpful when analyzing technology. But what is power?[5] We usually think of power in the *relational* sense—that is, "power over" others and as such is exercised through force, manipulation, and coercion—or in the *dispositional* sense—which refers to the capacity to bring about outcomes, that is, the "power to" do something. But power can also be *systemic*, which refers to the idea that power is also expressed through structures such as ideologies, laws, economic relations, families, etc. Finally, power can be *constitutive*, which refers to the ways individual subjects internalize relational and systemic power and turn it into norms and beliefs that affect behaviors and choice architectures. So can the multifaceted concept of power inform us as computer educators? Can it help us in the classroom better evaluate the social import of our field? The answer is yes!

---

[5] The following four-fold categorization is adapted from Sattarov [152] and Waelen [177].

Recall the argument above that in academic computing we should consider not only the algorithmic approaches used in AI, but also consider the vast infrastructure required to run it as well as the variety of social practices that are connected to its use. It is with these latter two aspects where power analysis can be fruitful. Crawford's superb *Atlas of AI* [45] illustrates the tremendous amount of human and natural resources required to develop and run AI-based technologies: resources that require not only dispositional power, but relational and systemic as well. For instance, usage of natural resources requires obtaining the financial capital that fuels dispositional power; but it also allows those with it to exert relational (i.e., coercive) power over large actors such as state governments competing to attract investment and small actors performing data tagging for penurious rates of pay. Systemic power is at play in the social practices of AI. Advantageous legal and intellectual property regimes for digital firms are part of their systemic power. But so is the ability of these firms to use the power of their platform to act as opinion leaders and shape the way individuals think about the social meaning of these technologies [87]. In other words, a key aspect of systemic power is its ideological expression: the capacity of firms (and of our academic discipline) to divert attention away from the inequality and injustice they are reproducing by making it appear right and just. As noted most famously by Foucault [116], this ideological expression of systemic power often becomes internalized by those subjected to it, thereby turning it into constitutive power.

The constitutive power potential of AI has already become a worry within legal scholarship under the label "algorithmic governance". The idea here is that individuals' choice environments are being excessively shaped by algorithmic systems. The selections made by our digital ecosystems, it is argued, "influence not only what we think about (agenda setting) but also how we think about it" [98]. Ominously, these worries about algorithmic governance were motivated by the much simpler machine learning systems of the previous decade. Such algorithms presented users with constrained sets of choices aimed at keeping them engaged. Yet by the end of the 2010s, this limited form of machine learning had given rise to a variety of worries about polarization, addiction, false news, information poisoning, echo chambers, and so on [141]. If this relatively straight-forward form of ranking AI has given us a world in which two neighbors could live in completely different epistemic worlds, then what fate awaits us in the next decade with generative AI, in which the human creator has been removed from the process and where the potential constitutive power of those using these systems is significantly magnified? In retrospect, both academic computing and more broadly, society itself, might have had an opportunity to address these potential problematic features of the previous generation of algorithmic ranking systems during the early years of their adoption. Prophetic early voices such as Sunstein [165] warned about these risks during the high noon of the dotcom era and advocated for a stronger regulatory environment, but notwithstanding some critical voices within the computing discipline, both it and its surrounding society were arguably [12] in the ideological grip of an overly-optimistic belief in the emancipatory power of networked computing.

Computing educators can play a role then in countering the constitutive power of such an ideology now that we are once again in the early years of another technological transformation. As argued by Giroux [77], "pedagogy is always political" and must be about more than skills acquisition or expanding coverage to a wider range of students. But, as noted by Shad and Yadav [157], the focus within academic computing education over the past 20 years has principally been on furthering the goals of computing and not on the more complex values of its teachers, its students, and their surrounding society. Indeed, computing practitioners must more often ask themselves questions such as: "is more computing really a good thing?" or "should this computing tool be built at all in a world riven by entrenched structural power asymmetries?" [182]. This type of critical questioning is becoming much more common within computing education. It can be seen in a wide variety of recent educational interventions within computing, such as critical digital literacy [143], computational action [167], critically conscious computing [106], equity pedagogy [117], critical computational literacy [112], culturally-responsive computing [153], counter-hegemonic computing [58], abolitionist computing [97], computational empowerment [93], liberatory computing [178], and justice-centered computing [109]. While this may seem a daunting list of approaches, there are perhaps more commonalities here than differences. In particular, they all advocate for a more critically reflexive perspective towards academic computing, which will be the focus of the next section.

## 4    Expanding the habitus of academic computing

> "What is characteristic of modernity is not an embracing of the new for its own sake, but the presumption of wholesale reflexivity which of course includes reflection upon the nature of reflection itself."
> — Anthony Giddens [76]

Disciplinary divisions of labor exist in all fields; this is also true of academic computing. Our training, our inclinations, our capabilities, our slow induction into the computational way of thinking—we might group these together and simply call them "dispositions"—all shape and guide us into a place in that division of labor. We've all had a student or colleague whose dispositions seem a "natural" fit for some type of computing practice, whether it be numeric methods, human factors, or algorithmic evaluation. As such, not every computer scientist is predisposed to taking an interest in educational research or in the normative aspects of their work. What are some of the dispositions of a computing academic interested in such topics? I'm not trying to make an essentialist claim: there are no necessary psychological, intellectual, or social characteristics for someone to be interested in computing education research and practice, nor do I wish to reify some type of ideal personality type. Nonetheless it doesn't seem implausible to posit the possibility that many of us interested in computer education (CEd) research or in computer ethics and society studies (CESS) share some dispositions.

As CEd and CESS researchers we have the predisposition, the training, the theoretic perspective, and the willingness to evaluate computing outcomes from outside of our "normal" disciplinary matrix. Using the terminology of Bourdieu [179], we inhabit a slightly different habitus from our computing colleagues who have no interest in these topics. Habitus is typically understood as the values, dispositions, and practices that agents gain, partly from their personal histories and partly from their membership in a social, cultural, or technical field [120]. These practices and ways of seeing the world operate somewhat unconsciously, and while it is by no means deterministic or immutable, a habitus predisposes one towards certain beliefs and actions. So what is the "normal" habitus for those in computing and how might those in CEd and CESS research differ? My thinking about this question was influenced by Bourdieu's description of reflexive practice [154] and the public sociology argument of Burawoy [27]. In his presidential address for the 2004 Annual Meeting of the American Sociological Association, Burawoy challenged his colleagues to move outside the university into the realm of activism and to engage in public discourses about society could or should be. Burawoy called this type of activity "Public Sociology" and presented it as one of a four-part division of intellectual labor in the field. His address inspired a substantial literature in response [5, 37]. So while computing and sociology are different disciplines (though I have previously argued that they both social sciences [redacted]), a disciplinary classification system can help us conceptualize academic computing's habitus.

In Table 1 (inspired by Burawoy [27]), I have classified the different aspects of academic computing labor based on the type of academic practice. I use the term "Instrumental" to capture the practical aspect of much of the labor that happens in academic computing disciplines; it is all about problem solving, constructing research programs, funding those programs, or applying this research. Most of us spend most of our time in the instrumental dimension. Indeed, without the activities of Professional Computing, there is no academic computing as it provides the scientific knowledge base for the discipline itself. Equally, Professional Computing is vitalized, legitimized, and funded thanks to the practical application of computing research outside the university in what I call Industrial Computing.

Table 1: Division of Academic Computing Practice

| Type of Practice | Internal Orientation | External Orientation |
|---|---|---|
| **Instrumental** | *Professional Computing*<br>    Research that defines theories,<br>    addresses questions, presents<br>    solutions | *Industrial Computing*<br>    Application of research, external funding,<br>    entrepreneurial monetization, filing patents,<br>    consulting, devising model curricula. |
| **Reflexive** | *Critical Computing*<br>    Internal debate within and<br>    between research programs,<br>    self-critique, critical<br>    research/theory | *Public Computing*<br>    Teaching, presenting for non-specialists,<br>    social media interaction, connecting<br>    discipline to concerns of wider world,<br>    political activism |

But academic computing has a neglected other dimension. We can think of it as a reflexive dimension because the analytic spotlight is turned upon ourselves as a discipline. Reflexive practice perhaps feels a little foreign to us in academic computing. "Normal" disciplinary work takes as a given the concepts, values, methods and presumptions of the field (Bourdieu refers to these collectively as *doxa* [48]). A great deal of our effort as computing educators is involved in habituating our students to the doxa of computer science. Reflexivity is a break from this unquestioned acceptance that we normally have in our field's doxa and introduces a perspective of doubt. In the somewhat enigmatic explanation of Bourdieu, reflexivity is "the systemic exploration of the unthought categories of thought which delimit the thinkable" [120]. Bourdieu and Burawoy argued that reflexivity can help to (partly) overcome some of the limitations under which any field operates by encouraging the interrogation of that field's doxa and habitus. In Table 1, this aspect is called Critical Computing. This willingness to engage in self-critique has sometimes been missing from academic computing. In the first decade of the 2000s we perhaps overly-focused our attention on self-promotion [157]. Arguing for the importance of computational thinking, pushing for the introduction of coding into P-12 education, and advocating for the expansion of computational approaches into other disciplines, all helped to preserve the discipline in the face of the harrowing enrollment crisis of the early 2000s [40]. But those efforts may have postponed the recognition that self-critique is an important constituent of any discipline.

But what exactly is Critical Computing? Isn't every form of research critical? Critical is certainly a word with many meanings and connotations. The idea of criticality embedded in the typology of Table 1 is referring to a general type of research paradigm known as Critical Theory in the social sciences [64, 23, 71] or Critical Research / Engagement Research in more practical fields such as Information Systems [135, 32] or Accounting [34, 44]. This paradigm refers generally to an overall strategy for conducting an inquiry in which a researcher's value orientation is at the forefront of the enterprise [80]. Instead of imagining research as the rigorous pursuit of objective explanations, this type of critical research is typically motivated by three goals. First, to *gain a* *deeper insight* into a phenomenon/institution/belief in the human or social or technical world by making use of social theory with a critical orientation (for instance, by Habermas, Bourdieu, Foucault, Freire, Butler, Kristeva, Crenshaw, Spivak, Haraway, Mbembe). Second, building upon the insight gained, *engage in diagnosis and critique* of the phenomenon. This may involve identifying the ideological bases or contradictions of beliefs/practices that make up the phenomenon. This type of critique requires the researcher to take an explicit value position and advocate for explicit moral principles that are used to judge and critique. These principles are typically one or more of equality, justice, democracy, capabilities, emancipation, or sustainability. This explicit value position is key to the third goal: *suggesting improvements* or developing an account of viable alternatives. That is, the transformation of an unequal, unjust, undemocratic, unsustainable, etc phenomenon is the ultimate, "real utopian" goal of critical practice [185]. Not everyone working in the reflexive dimension needs to be involved in all three goals, but as a disciplinary field, it is essential that more researchers engage in one or more of them. Happily, in the HCI, computer education, and Computing for the Social Good subfields, there is a growing set of scholars already working toward these critical goals (many of which have been referenced in the preceding pages).

Finally, the other half of reflexivity is the dialogue between ourselves and those outside the discipline, which in Table 1 is called Public Computing[6]. A dialogue is not a lecture at someone; it is a process of mutual education, a dialectic in which we also learn from the audience. This dialogue can be with ourselves (Critical Computing), but it also can be with the outside world (Public Computing). That is, all the activities in the reflexive dimension must be dialogical as there is no clear and well-accepted way of adjudicating the knowledge claims in this dimension. This doesn't mean it is inferior to instrumental practice; instead, it should be understood as a necessary part of the discipline that runs on different principles.

This can be seen if we look at the two types of academic practice that interacts with the world outside of our discipline (i.e., those in the External Orientation column). Both interact with the public but differ completely in their motivation. In Industrial Computing, the public is an object not a subject: something to be manipulated, an instrument for achieving the goals of the researcher (or discipline). That is, its goal is not the explicit betterment of some public; rather it is focused on the self-interested needs of the individual researcher, their graduate students, or the discipline as a whole. But in Public Computing, the computer academic is engaged in a dialogue with others outside of computing. It treats the public as a subject, that is, as something to be valued not in terms of what it can do for us, but because the public has an intrinsic value. In Kantian ethical language, the public is more than just a means, it is an end in itself. This is why political activism is present as part of Public Computing. As emphasized in the second section of this paper, much of what is produced, researched, and taught in computing is ultimately a type of political action, albeit sometimes hidden away. But an explicitly public computing is not content with an implicit political computing; instead, it is willing to embrace the need for political activism in order to improve the world in line with the goals of critical computing.

Another way of comprehending the key differences between the four quadrants is by looking at the different modalities within the practices, which can be seen in Table 2. Each of these quadrants differs substantially in their cognitive practices: not only in terms of the nature of their knowledge and how truth claims are legitimated, but also in terms of their accountability, their relationship to other disciplines, and their motivation. It also illustrates why connections between the four different types of academic labor can be difficult and why the perception that some CEd and CESS researchers are inhabiting a different epistemological frame than the rest of our computing colleagues [147].

---

[6] It should be noted that the "public" and its related cousin the "public good" are very much an area of contestation. Contemporary societies are not characterized by unity but by competing groups dominated in turn by hierarchies of power. So it is important for choices to be made about "which public" and "whose public good". This question is raised again in the next section. A more comprehensive examination of this question can be found in [redacted].

Table 2: Modalities of Academic Computing Practice

| Type of Practice | Internal Orientation | External Orientation |
|---|---|---|
| **Instrumental** | *Professional Computing* | *Industrial Computing* |
| Epistemology | Theoretical / Empirical | Practical / Pragmatic |
| Legitimacy | Scientific / Mathematic Norms | Effectiveness / Performance Metrics |
| Accountability | Peers | Clients |
| Motivation | Self-Interest | Self-Interest |
| Interdisciplinarity | Cross-Disciplinary Borrowing | Stakeholder Domains |
| **Reflexive** | *Critical Computing* | *Public Computing* |
| Epistemology | Normative | Dialogical |
| Legitimacy | Moral Vision | Politics |
| Accountability | Critical Intellectuals | Publics |
| Motivation | Internal Debate/Critique | Social Responsibility |
| Interdisciplinarity | Transdisciplinary Infusion | Multidisciplinary Collaboration |

 Of course, practices in the real-world of academic experience may straddle these ideal types, but this categorization might help us situate particular ethical concerns and education research as a whole. Given this characterization of the expanded overall habitus for academic computing, it seems plausible to assert that both ethical/political evaluation and computing education research naturally inhabit the reflexive dimension of academic computing. Education is by its very nature a public practice. It is much more than the provision of knowledge and skills; it is a collaborative process in which identities are constructed and agency—the ability to act and think in the world—is enhanced for both teacher and student.

In this formulation of the reflexive dimension of computing, critique and public responsibility are the key motivations. It is my contention that we need to include more of the reflexive dimension in our disciplinary work. However, it is worth acknowledging that doing so would position the reflexive computer academic in a somewhat fraught attitude towards her/his/their disciplinary home. To make an argument for how we might successfully embrace this dimension, the next section is going to resurrect the archetype of the public intellectual and advocate for a new role within academic computing: the Public Computer Intellectual.

## 5    The public computer intellectual (PCI)

"There is no such thing as a private intellectual, since the moment you set down words and then publish them you have entered the public world."

— Edward Said [149]

All humans are intellectual but not all humans have in society the function of intellectuals Gramsci famously stated in his *Prison Notebooks*, which contains an early and influential account of intellectuals and their role in society [155]. Intellectuals for Gramsci include scholars, artists, and organizers of culture. He distinguished two types of intellectual: traditional intellectuals—old-fashioned generalist scholars who are integrated into a historical tradition and who claim to be independent of politics—and organic intellectuals, who grow out of the needs of a social class and thus are more directly tied into the economic structure of their society.

Since that time, the ideal of public intellectuals has occasionally been resurrected [94, 144, 161]. The traditional intellectual in the Gramsci sense has, since the late twentieth century, become reliant on university employment and thus can no longer be generalists but must instead be specialists. This universitization of intellectual life has tended to remove a lot of the "public" from "intellectual" work. Nevertheless, what remains of the public intellectual archetype is, despite the specialization imposed by academic life, still "marked by the ability to draw on a wide range of disciplinary insights, frequently versed in fields and literature beyond of their academic home" [49]. These more recent works on public intellectuals has tended to see them in a more favorable light than Gramsci; public intellectuals today are much more likely to publicly engage in debate and advocation on behalf of the less privileged. In the North American context, some exemplars from the past twenty to thirty years would include Camile Paglia, Edward Said, Noam Chomsky, Margaret Atwood, Christopher Hitchens, Cornell West, Angela Davis, and Naomi Klein.

Of these, Edward Said—author of *Orientalism*, classical music critic for *The Nation*, and outspoken defender of Palestinian rights—wrote about intellectuals in a manner quite relevant to this paper's idea of a Public Computer Intellectual (PCI). For Said, the intellectual "is neither a pacifier nor a consensus builder, but someone whose whole being is stalked on a critical sense, a sense of being unwilling to accept easy formulas, or ready-made cliches … [and who is] actively willing to say so in public" [149]. The vocation of the public intellectual is thus all about "maintaining a state of constant alertness, of a perpetual unwillingness to let half-truths or received ideas steer one along." For Said, this predisposition is tied to two related forms of reading / observation which he called "reception" and "resistance" [148]. Reception, for Said, refers to a willingness to carefully unearth the messages being communicated by some cultural artefact; resistance is a willingness to be publicly uncomfortable with those messages when they conflict with the good of the less powerful. The task of an intellectual is thus "to be both insider and outsider to the circulating ideas and values that are at issue in our society" [148].

This, then, is what computing needs in an age of AI crisis: a perspective based on disciplinary close-reading (reception) that interrogates the value premises of the academic computing field and thereby takes a stand for the needs of the public instead of the needs of the profession (resistance). This requires recognizing that there is more than one public. We indeed live in a society fractured into divided, sometimes hostile, publics. Public Computer Intellectuals thus need to decide which publics' interest they wish to promote. The entire instrumental dimension of academic computing practice (from Table 1 and 2), which takes up the lion's share of our disciplinary attention, arguably principally benefits the already privileged. Perhaps, then, following Piven [142], we ought instead to "strive to address the public and political problems of people at the lower end of the many hierarchies that define our society." Given computing's close historical relationship with power—indeed the device from its very beginnings to the current day has been a mechanism for expressing and extending the dispositional, relational, and systemic power of the already powerful [192]—it arguably makes sense in terms of fairness for a public and critical computing to ally itself more with addressing the problems and needs of the less powerful.

Consequently, a PCI should be characterized by a willingness to be a critic and dissident within the discipline and thus to inhabit both Critical Computing *and* Public Computing. There are already exemplars in this sense who can inspire us. Here I will mention Abeba Birhane, Joy Buolamwini, Kate Crawford, Amy Ko, Timnit Gebru, Mark Guzdial, Arvind Narayan, and Jonathan Sadowski as computing academics who are also Public Computing Intellectuals. Abeba Birhane (Senior Advisor in AI Accountability at Mozilla Foundation as well as Assistant Professor in Computer Science at Trinity College, Dublin) has documented the discriminatory malfeasance which lurks within many AI models, products, and research [16, 17, 18]. What turns her into a PCI is that she resides not only in the Critical Computing quadrant but also within the Public Computing one. She tirelessly brings to light the many politically problematic features of contemporary machine learning research, not only in academic publications but on Twitter as well. Timnit Gebru (formerly of Google), Kate Crawford (Research Professor at USC Annenberg), and Arvind Narayan (Professor of Computer Science at Princeton) have also been critical voices within computing; part of the way they have expressed their public aspects have been through media appearances, through blog posts, through articles in the public press, and through their work in organizations devoted to the public interest (the Distributed AI Research Institute, the AI Now Institute, and the Princeton Center for Information Technology Policy respectively). Jonathan Sadowski, Senior Research Fellow in the Department of Human-Centered Computing at Monash University, focuses on the political economy of computing. His public-facing books and almost 100 articles in non-academic sources are situated in the critical paradigm and focus on how computing's close connection to technological capitalism influence the design and use of computing. He also co-hosts the very popular podcast *This Machine Kills*.

Joy Buolamwini may be the ultimate PCI. Co-author of the highly-cited "Gender Shades" paper [25] as a Masters student, she started the public-facing Algorithmic Justice League as a student, presented multi-media poetry about the intersectional injustice of AI systems, appears frequently in US news outlets, and spoke at official government hearings for the EU and the US Congress. Buolamwini also has been involved in practical political and legal activism for less-advantaged New York tenants fighting mandatory dataification imposed by their landlords. Her public message about computing over the past five years has been that "the privileged few were designing for the many with little regard for the harmful impact of their creations" [26]. As noted in her recent autobiography *Unmasking AI*, her tireless public work not only gained her a PhD but also a "seat at the table" meeting with President Biden.

Computing education research also has its PCIs. Amy Ko, currently editor of *ACM Transactions on Computing Education*, has been a public advocate for broadening research paradigms in computing as well as encouraging the adoption of a justice lens within our subfield. With a series of student collaborators, Ko's work has focused on discrimination and bias in computer systems [106, 180] and has brought that work to a wider public on Twitter and Medium. Mark Guzdial is similarly engaged in an active dialogue with both the profession and the public about the nature of computing education. His perspective may at first glance not seem especially political. But his tireless advocacy—in both academic articles and in Twitter, blog posts, and public-facing magazine articles—on behalf of his belief that CS needs to change "so it serves the needs of our students and society" [81] fits both reflexive quadrants of Table 1. In all of these cases, these PCIs are commentating in an accessible manner on technical material that would otherwise remain out of sight for those outside of academic computing. Following Table 2, their technical critique from within is an example of accountability to a public and takes for its legitimacy a moral vision of what education and society should and should not be.

Of course, not every academic wishes to have a public presence. What if one doesn't want to voice their opinion or research on X, Medium, YouTube, or the *New York Times*? Indeed, being outspoken public critics appears to have incurred a real emotional or professional cost to some of the above exemplars. Furthermore, public forms of engagement and activism that are outside of academic conferences or journals can be considered non-legitimate which it comes to academic promotion decisions. This is especially true for less-established scholars or scholars outside of elite universities or without connections to powerful tech companies, who might not be able to attract sufficient viewership on these public platforms to impress a skeptical tenure or promotion board.

There are some mechanisms that might help in this regard. One practical mechanism that might allow for an expanded expression of the reflexive dimension in academic computing would be to institute an expectation that aspects of the reflexive dimension be part of the peer review process in computing. This was the principle behind the workshop that Sturdee *et al* ran at CHI'2021: "to explore and establish the principle that the potential negative consequences of [computing] research should be questioned, critiqued, and discussed as part of the publication and peer review process" [163]. Recognizing the value of public dialogue in the promotion process would also be an important way to institutionalize the recognition of the reflexive dimension. This paper is also partly intended to convince more in our discipline to accept a wider-range of academic practice as legitimate scholarship.

But the key point I wish to make with the PCI construct is that these public intellectuals within computing help to exemplify many of the virtues of the reflexive dimension of academic computing practice. It is not meant to be a demand that more of us should talk to Congress, host a podcast, or be featured in *Time Magazine*. Yes, some of these exemplars were able to engage with publicly-recognized outlets, but that isn't why they are important. What matters is that each of them inhabits both halves of the reflexive dimension in their academic practice; they use criticality in all the senses described earlier, not for career progress, but as a way to defend a public's interest. This is a way of being, then, a way of practicing academic computing that we can all inhabit a bit more in our research and in our teaching.

As such, we do not have to be actively engaged in the public's eye in order to inhabit the virtues of the PCI in our teaching and research. That is, we can all be an *(Almost) Public Computer Intellectual*—an (A)PCI —, whether in the classroom or in academic research. An (A)PCI can still be an advocate in the classroom, not for computing itself (that's the job of the ACM/IEEE, university program advisors, or industrial spokespeople), but for those who often have to bear the brunt of computing advances: the exploited without voice in our field, mutely confounded by opaque algorithmic systems protected by a fog of complex mathematics and the ideology of inevitable technological progress. This means that as (A)PCI's we should devote more of our academic attention in computing, both as educators and as researchers, to the needs of the poor, to racial minorities, to women more than men, to the marginalized without legal residence, and to the world outside of the wealthier developed core. As advocated by critical theory [23, 71], we want to identify the potential for emancipation in the outputs and thinking processes of academic computing; we also want to bring to light the lived reality of actual repression that is a consequence of the way our discipline is integrated into the public economy of late capitalism.

Most of the PCIs highlighted above are disciplinary experts in the fields they are publicly critiquing. But the reflexive dimension is available to all. Computer education researchers may be also well suited to being A(PCI)s. Partly this is due to disposition (i.e., by the fact that the habitus education researchers inhabit already embraces both quadrants of the reflexive dimension) and partly due to the emancipatory nature of education itself: that is, to education's concern for justice, equality, and capabilities in educational experiences. Educational pedagogy has always been a "discourse of both critique and possibility" [77], a conversation not just among educators, but a dialogue between educators, students, and different publics. As mentioned at the end of this paper's section on Educational Approaches to the AI Crisis, there has been an explosion of interest in adopting both critical approaches and public-facing political activism in how computing is taught both in P-12 and in higher education. Arguably—to this author at least—recent trends in non-computing pedagogical research seems to be also encouraging a more political education [e.g., 114, 125, 103, 29]. As noted by Giroux, political education is different than a politicizing education. In the latter, we tell the students to think as we do; in the former, we teach them "by example the importance of taking a stand" [78]. This is why all computing educators could become (A)PCIs, even if the dimensions of their public is circumscribed by the walls of the lecture hall.

One final qualification needs to be made about the ideal of a PCI or a A(PCI). The focus on individual intellectuals certainly appears to be an individualistic approach to a set of problems which requires collaborative solutions. This is an issue also noted by Bourdieu in one of his last essays. It was titled "For a Scholarship with Commitment" [21] and it argued that educators as intellectuals are essential to the social struggles for a better world. But he also emphasized that what is required is a "collective intellectual" not a whole host of individual intellectuals. That is, changes—whether they be in an academic discipline or in the wider society—won't happen without a tremendous amount of solidarity being built. How to create solidarity in contemporary society is a complex topic and a daunting task [90], far beyond the scope of this paper. But the academic solidarity needed to create a collective intellectual in academic computing seems a more manageable possibility. It will for sure require more reflexive practice. But how do we create more disciplinary capabilities in the reflexive dimension? One mechanism would be to make productive research connections with colleagues in other disciplines who are already analyzing computing in the reflexive dimension. It will also require a change in the pedagogy of computer scientists, especially in graduate school. Recall from the previous section's discussion of critical computing that gaining a deeper insight into the social outcomes of computerization can be abetted by making use of already-established critical theories. For this reason, it would make sense for computing students to read and debate some of these theories. Currently, many CS programs have an external philosophic ethics course [66]. This could be replaced by a social theory course offered by most sociology departments. For programs with a ethics course taught in-house by computing faculty, such courses could focus much less on ethical theories and much more on exposing students to the analytic riches (and problems) of critical theory. Such an exposure would be especially important at the PhD level. It is interesting that even very professional and practical disciplines such as accounting [34], information systems [135], and the health sciences [139] have been embracing and integrating critical theory approaches into their graduate school practices. Another mechanism for constructing a stronger reflexive dimension is through the hiring practices of the discipline. As this author has previously argued [redacted], there are many PhDs being completed in disciplines outside of computing that are often explicitly focused on social and psychological consequences of computing; hiring committees could be encouraged to be more willing to countenance expanding the range of expertise in their departments by hiring such individuals. Finally, the PCI construct from the previous section valorized a variety of outstanding individuals and their accomplishments. But it is the (A)PCIs that are the true vector for change. Engaging in "minimal utopianism in the classroom" [20] can be one small step in constructing a collective intellectual that includes both students and teachers. Encouraging collective experiences with the world outside the university—such as those described by Lachney and Yadav [108] and Rivera and Su [145]—are another way of fighting against the solitary subjectivities created by 21st century digital capitalism.

## 6    Conclusion

"There is nothing more important for everyone, but particularly for young people, than being attentive to the signs that something different from what is happening might happen."
— Alain Badiou [9]

This essay has argued that the public narrative around the AI crisis is worth attending to because it can help us see how vital it is to expand the computing discipline's focus beyond technological topics so as to see the wider background affected by its work. In this regard, the lenses of economics, sociology, political science, and other social sciences are necessary. In particular, power as an analytic category is especially valuable for understanding the relationship between computing and the rest of society. Such an approach aligns well with contemporary educational theory since both share an explicit normative goal of emancipation and empowerment. This paper then presented an expanded categorization of academic computing that legitimates both internal critique and a broader concern for the public interest as constitutive aspects of academic computing practice. Finally, Public Computer Intellectuals and, just as important, the (Almost) Public Computer Intellectual, were presented as an example of what such an expansion could look like.

Computing today is a provocative topic. Citizens young and old are very much engaged in debates around the various changing manifestations of computing technology. These debates are not around the technicalities of the algorithms; rather, they are about ramifications, often initially on the personal or familial, but also on the wider social world. Are my teenagers' preoccupation with visual spectacle on Instagram and TikTok making them superficial? Has a reliance on quick and easy web searches made us progressively unwilling to countenance prolonged argumentation? Is online pornography contributing to a decline in libido? If I spend more and more time online exposed to ideas congruent with my own, will this make my opinions more uncompromising and extreme? Will ChatGPT affect my job? Will it make my children less able to write and think on their own? Is it fair that AI image generators profit privately from the consumption of billions of public images without the consent of their owners?

These are the debates of our time, much as class struggle was to the 1920s and 1930s, the struggle against fascism and communism was to the 1940s and 1950s, decolonization and mass consumerism was to the 1960s and 1970s, and the transformation of life around the exigencies of globalized finance was to the 1980s and 1990s. Public debate about these issues in decades past was the purview of experts in the media, in independent thinktanks, or in the policy departments of universities. But now it is our—that is, academic computing's— moment in the sun. We have a duty and responsibility to add our voice, our expertise, to these debates.

But to do so requires more than just a willingness to participate in these debates. It requires a recalibration of computer science's intellectual orientation to be more focused on the present and future world outside academia. We should no longer think of our discipline (and ourselves as computer education practitioners) as being only a technical one akin to our cousins in the natural and engineering sciences, operating according to unproblematic natural laws. This would require a willingness to teach a wider range of disciplinary material to computing students; this is especially important within graduate programs. As well, within computing education research we need to continue expanding beyond the problem of how to improve student learning of the technical aspects of computing. We need to recognize that encouraging a critical stance towards the discipline is just as important. Exposing students to critical social theories—either as a supplement or even as a replacement for teaching macro-ethical theories—can be one straightforward way of contributing to that goal that doesn't require wholesale changes to our curriculum.

In 2022, Judith Butler, famous for rebarbative texts for theory specialists such as *Gender Trouble* (1990), *Bodies that Matter* (1993), and *Giving an Account of Oneself* (2005), wrote a beautifully profound and accessible piece for *Daedalus* on the public future of her discipline. In it, she wrote that the question of the future is "the predominate problem for the humanities" [28]. This paper has argued that the future should also be the predominate problem for the computing discipline. Moving forward in a soon-to-be world where the AI crisis is transforming from a what-might-be to a what-now-is, the need for an institutionalized acceptance of the reflexive dimension within computing will be increasingly needed.

This essay began with a quote from Gramsci on crisis, and how the period between the decline of an old way of life and before the ascendency of a new one is a time of both risk and opportunity. Gramsci felt public intellectuals could play a key role during such times in helping society navigate a path to a less unpleasant future. The AI world of the future is one of unprecedented risk. Within academic computing we have both a tremendous responsibility to the future as well as an opportunity to effect a transformation in how we think about computation and teach its principles. There are of course many fetters—such as hiring practices, the publishing expectations of tenure boards, curricular inertia, and the increasing demand by governments that research be more focused on the needs of industry—that act as impediments to achieving such changes. Yet constraint is the condition of possibility, as Butler noted more than three decades ago in *Gender Trouble* in regards to what seemed the intractable problem of transforming social attitudes towards gender [quoted in 59].

That is, the very fact that there are difficult constraints on free action can sometimes open up the best possibilities for transformation as the constraints provide compelling motivation for thinking (and acting) in different ways. This essay has argued that embracing the critical and public practices in the reflexive dimension is one possibility for thinking and acting in new (or renewed) ways. Both Public Computer Intellectuals and the (Almost) Public Computer Intellectuals within CEd and CESS research can play a role in helping to achieve this transformation.

## Acknowledgements

I was very fortunate to have this paper markedly improved by the excellent ideas and suggestions of three anonymous reviewers as well as my colleagues Janet Miller, Marc Schroeder and Marty Clark.